\documentclass[preprint,aps,nobibnotes,tightenlines,10pt]{revtex4}%
\usepackage{amsfonts}
\usepackage{amsmath}
\usepackage{amssymb}
\usepackage{graphicx}%
\begin{document}
\preprint{UATP/04-07}
\title{Configurational States and Their Characterization in the Energy Landscape}
\author{P. D. Gujrati and F. Semerianov}
\affiliation{The Department of Physics, The Department of Polymer\ Science, The University
of\ Akron, Akron, Ohio 44325}
\date{\today }

\pacs{PACS number}

\begin{abstract}
Configurational states that are to be associated, according to Goldstein, with
the basins in the potential energy landscape cannot be characterized by any
particular basin identifier such as the basin minima, the lowest barrier, the
most probable energy barrier, etc. since the basin free energy turns out to be
independent of the energies of these identifiers. Thus, our analysis utilizes
basin free energies to characterize configurational states. When the basin
identifier energies are monotonic, we can express the equilibrium basin free
energy as a function of an equilibrium basin identifier energy, as we explain,
but it is not necessarily unique.

\end{abstract}
\maketitle

It is well known that most supercooled liquids (SCL) become viscous when their
configurational entropy becomes negligible as they are cooled, provided the
corresponding crystal is not allowed to nucleate.\ Our current understanding
of glassy behavior is still far from complete, even after many decades of
continuous investigation. In order to better understand the flow properties of
viscous fluids, Goldstein proposed the potential energy landscape (PEL)
picture using \emph{classical\ canonical ensemble }\cite{Goldstein,Goldstein1}
to qualitatively discuss an interesting and sufficiently tractable scheme to
study SCL and the glassy state by drawing attention to the potential energy
basins that are expected to play a pivotal role in the thermodynamics of
viscous fluids at \emph{low temperatures}. The PE surface contains many local
potential energy basins, each possessing a \emph{basin minimum} (BM). The
study paved the way to exploit PEL to study \emph{stationary} SCL by the use
of the partition function (PF) formalism. The stationary state is obtained
under infinitely slow cooling of the disordered equilibrium liquid below the
melting temperature $T_{\text{M}}.$ The basins have other points of
interest\ such as the lowest energy barrier (LB) or the most probable energy
barriers, i.e. the barriers with the highest population (MPB) that most
probably play an important role in crossover dynamics or diffusion, as the
temperature is raised. We will collectively call these basin points of
interest basin identifiers. Goldstein provided a qualitative description of
the nature of the resulting landscape created by the distribution of basins
and their minima, and drew attention to the idea of \emph{configurational
states}, which should be associated with individual basins and should change
with the temperature $T.$ Instead of being determined by the particular
topology of the basin, the configurational state of the system associated with
that particular basin, according to Goldstein, is uniquely specified only by
the potential energy $\mathcal{E}$ of the basin minimum$.$\ This will
certainly make sense if the equilibrium value $\overline{\mathcal{E}%
}=\mathcal{E}(T)$ of $\mathcal{E}$\ is in one-to-one correspondence with
(i.e., is a \emph{monotonic} function of) the temperature $T.$ We wish to
emphasize that throughout this work, we are only interested in non-negative
temperatures.\qquad\qquad

As we will see below,\ there is no thermodynamic requirement for
$\mathcal{E}(T)$\ to be monotonic. Even if $\mathcal{E}(T)$ is monotonic, is
expressing thermodynamic quantities as a function of $\mathcal{E}(T)$ over a
wide range of temperature an oversimplification, since it makes the shape and
topology of the basins irrelevant? For example, if we consider a particular
basin, then its free energy changes with $T$, but the basin minimum energy
remains unaffected by $T.$ Similarly, the free energy also changes as the
system moves from one basin to another at the same temperature. The BM energy
$\mathcal{E}$ has no guarantee to remain the same. This change in
$\mathcal{E}$\ is certainly not going to be reflected in $\mathcal{E}(T)$.
Thus, we expect not only the equilibrium $\mathcal{E}(T)$ but all equilibrium
quantities\ to have separate intrabasin and interbasin contributions. Indeed,
this happens as we will see here [see (\ref{FinalBMEnergy2}) below].$\ $We
find that the equilibrium value of every thermodynamic quantity including
$\overline{\mathcal{E}}$ is a function not only of $T,$ but also of the basin
free energy. Only at low temperatures do we expect the latter dependendence on
the free energy to become weak as assumed by Goldstein. In that case, the use
of $\mathcal{E}$\ may lead to erroneous description if care is not exercised.
What about other points like LB and MPB in the basins? Will they play a more
important role at higher temperatures? Is there a thermodynamic basis for
their importance?

It is obvious that the shape and topology uniquely determine the free energy
of the basins, which then uniquely determine the thermodynamics of the viscous
fluid. Should we not describe configurational states by specifying the basin
free energy? The latter is certainly monotonic in $T.$ We note in this regard
that, based on an analogy with spin glasses, the use of basin free energy in
place of the BM energy, as originally advocated by Goldstein, has been
proposed \cite{Parisi,Parisi1,Coluzzi,Coniglio}. An alternate approach is to
use the free energy landscape \cite{Anderson} and has been applied, for
example, to dense hard sphere system \cite{DasGupta} to study glassy behavior.
Here, we are only interested in the potential energy landscape. The use of PEL
has now become common in many disparate fields like glasses, proteins and
clusters \cite{Wales}, and has established itself as an important
thermodynamic approach in theoretical physics. Thus, it is highly desirable to
answer the above questions. It is this desire that has motivated this investigation.

We briefly review the basics of the PF formulation in statistical mechanics
\cite{Landau} as some of the steps in the evaluation of the PF are going to be
crucial in our discussion here. The PF $Z(T)$ is a sum of the Boltzmann weight
$\exp(-\beta E_{n})$ for each configuration of energy $E_{n}$
\begin{equation}
Z(T)\equiv\sum_{n}e^{-\beta E_{n}}; \label{ConPF}%
\end{equation}
here $n$ indexes the configurations of the system. However, as we can change
the value of $n$ associated with a configuration at will by relabeling it, the
index$\ n$ is \emph{not} unique and, therefore, its possible values for
equilibrium configurations are not very useful for thermodynamic description.
Therefore, we need to relate $n$ to a unique thermodynamic property of all
configurations, and not just the equilibrium configurations. The property must
be such that its equilibrium value is a \emph{monotonic} function of $T,$ so
that it can be used in place of $T$ if we so wish$.$ It is not hard to
identify such a property. Since each configuration has a unique energy$,$%
\ which is a thermodynamic quantity, we can use the energy $E_{n}$ in place of
$n$ as the summation index. As the energy of interactions $E_{n}$ is coupled
to the inverse temperature $\beta,$ see (\ref{ConPF})$,$ the equilibrium
energy%
\[
\overline{E}\equiv E(T)\equiv\sum_{n}E_{n}e^{-\beta E_{n}}/Z(T)
\]
is a monotonic function of $T$ because of a non-negative heat capacity:
$\overline{E}$ and $T$\ are in a one-to-one correspondence. Thus, the first
step is to rewrite the PF in (\ref{ConPF})%

\begin{equation}
Z(T)\equiv\sum_{E}W(E)e^{-\beta E}, \label{StandardPF}%
\end{equation}
where $W(E)$ represents the number of configurations of energy $E$ and defines
the microcanonical entropy $S(E)\equiv\ln$ $W(E).$ The sum in $Z(T)$ at a
fixed $T$ is over all allowed energies $E$ and, by construction, each summand
is also a function of $E.$ These energies and, therefore, $S(E)$ do not depend
on $T$. The second step in the evaluation is to recognize that the value of
$Z(T)$ for a macroscopic system, which is what we consider here, is determined
by the dominant term in the sum. The dominant term corresponds to the
equilibrium energy $\overline{E}$ at which the summand in (\ref{StandardPF})
is maximum. Thus, $Z(T)\cong W[\overline{E}]e^{-\beta\overline{E}}.$\ It is
the presence of $W(E)$\ in (\ref{StandardPF}) that makes the summand strongly
peaked at $\overline{E}.$ The determination of $\overline{E}$\ for a
macroscopic system is simplified by noting that $E$ is almost a continuous
variable for a macroscopic system. In terms of $S(E),$ $\overline{E}$ is given
by the location of the minimum of the free energy function $F(T,E)=E-TS(E)$ at
fixed $T.$ In equilibrium, the canonical entropy $S(T)\equiv S(\overline{E})$
and free energy $F(T)\equiv F(T,\overline{E})=E(T)-TS(T)$\ become functions
only of $T.$ The conditions for the minimum are $[\partial F(T,E)/\partial
E]_{T}=0,$ and $[\partial^{2}F(T,E)/\partial E^{2}]_{T}>0$ leading to
\begin{equation}
\left[  \partial S(E)/\partial E\right]  _{E=E(T)}=\beta,\text{ }\partial
S(T)/\partial T>0, \label{EquilibriumE}%
\end{equation}
which are always satisfied in equilibrium.

At a given temperature $T,$ only those configurations that have the energy
$E=\overline{E}$ (or within a narrow width around it, depending on the heat
capacity; we will neglect this width here) determine the thermodynamics
through the entropy $S(T)$. All energies other than $\overline{E}$ and,
therefore, all configurations not included in $W(\overline{E})$ are irrelevant
at $T.$ Thus, thermodynamics is highly \emph{selective}. We will call
configurations with $E=\overline{E}$ equilibrium configurations to distinguish
them from all configurations. All thermodynamic quantities related to
equilibrium configurations will be similarly called equilibrium quantities to
distinguish them from quantities related to general configurations that may
not be equilibrium ones. In the following, we will make this distinction carefully.

As said above, the equilibrium values of $E$ are strongly peaked about
$\overline{E};$ all these configurations have the same minimum free energy
$F(T).$ These configurations belong to different basins with different BM
energies. Therefore, $\overline{E}$ at a given temperature must be independent
of $\mathcal{E}.$ However, there is no guarantee that the the equilibrium
values of $\mathcal{E}$ are also strongly peaked about $\overline{\mathcal{E}%
}.$ In other words, there is requirement that equilibrium configurations all
belong to basins having their minima that are strongly peaked about
$\overline{\mathcal{E}}.$ It is certainly possible for equilibrium
configurations to belong to different basins that have their minima at
different energies, but still correspond to the same free energy $F(T)$
because of the shape of the basins. In this case, it does not seem reasonable
to classify equilibrium configurations only by the minima of the basins they
belong to. We must also use the basin free energy in addition to the BM
energies for this purpose as discussed above. However, it appears highly
plausible that it can be done at low temperatures as originally proposed by
Goldstein \cite{Goldstein,Goldstein1} since the basin free energies are close
to the BM energies in values.

In his analysis, Goldstein has considered basin minima as the basin
identifier, and has listed two conjectures that were common in the field
\cite{Goldstein1} at the time: the basin PF $z_{\text{b}}(T)$ is (i)
independent of the basin's minimum energy $\mathcal{E},$ and (ii) insensitive
to the basins being explored. Utilizing these assumptions, Goldstein has
expressed the PF as a product \cite{Goldstein1}\ of the basin and BM
PF$^{\prime}$s\ \ \ \ \ \ \ \ \ \ \ \ \ \ \
\begin{equation}
Z(T)=z_{\text{b}}(T)Z_{\text{BM}}(T);\label{GoldsteinPart}%
\end{equation}
here $z_{\text{b}}$ for a given basin is defined by considering shifted
energies $E-\mathcal{E}$ with respect to the minimum energy $\mathcal{E}$ of
that basin; see also Schulz \cite{Schulz}. Goldstein has emphasized that basin
anharmonicity or the curvature at its minimum \cite{Angell} may be very
important. These are included in $z_{\text{b}},$ so that it is determined by
the entire basin topology$.$ According to Goldstein, all equilibrium basins
have the same equilibrium basin free energy $f_{\text{b}}(T)\equiv-T\ln$
$z_{\text{b}}$. The BM-PF is defined \cite{Goldstein1,Schulz} as%
\begin{equation}
Z_{\text{BM}}(T)=\mathbf{\sum_{\mathcal{E}}}N_{\text{BM}}(\mathcal{E}%
)e^{-\beta\mathcal{E}}.\label{GoldsteinConfPF}%
\end{equation}
Here, $N_{\text{BM}}(\mathcal{E})$ represents the number of basins whose BM
are at energy $\mathcal{E}.$ The equilibrium BM energy $\overline{\mathcal{E}%
}=\mathcal{E}(T)$\ is the value of $\mathcal{E}$\ at which the summand in
(\ref{GoldsteinConfPF}) is maximum. The conditions for the maximum in terms of
the BM entropy $S_{\text{BM}}(\mathcal{E})\equiv\ln N_{\text{BM}}%
(\mathcal{E})$ are given by
\begin{equation}
\left[  \partial S_{\text{BM}}(\mathcal{E})/\partial\mathcal{E}\right]
_{\mathcal{E}=\overline{\mathcal{E}}\ }=\beta,\text{ }\partial\overline
{\mathcal{E}}/\partial T>0,\label{Goldstein1}%
\end{equation}
which are the standard conditions of equilibrium; compare with
(\ref{EquilibriumE}). It is clear that the BM description proposed by
Goldstein ensures that $\overline{\mathcal{E}}$\emph{ is a monotonic
increasing function of }$T.$ Since his approximation is expected to be good at
low temperatures, we expect $\overline{\mathcal{E}}$ to monotonic increasing
there. But we will see below that it need not be true at all temperatures,
thereby limiting the usefulness of the BM-description to which we now turn.

\textbf{BM Picture.} We take BM$^{\prime}$s as basin identifiers, but the
discussion is easily extended to LB$^{\prime}$s or MPB$^{\prime}$s, or other
basin identifiers. The PEL is a union of disjoint basins. A basin\ is indexed
by $j$, and the lowest and highest basin energies are denoted by
$\mathcal{E}_{j}$, and $\mathcal{E}_{j}^{^{\prime}},$ respectively, so that
the basin does not exist outside the energy range $\Delta_{j}E\equiv$
$(\mathcal{E}_{j},\mathcal{E}_{j}^{^{\prime}})$. \ Since PEL\ does not depend
on $T$, $\mathcal{E}_{j}$ and $\mathcal{E}_{j}^{^{\prime}}$\ do not depend on
$T.$ Let $W_{j}(E)$ ($E\in\Delta_{j}E$) represent the number of distinct
configurations of energy $E$ in the $j$-th basin and introduce the entropy
$S_{j}(E)\equiv\ln W_{j}(E)$. It is obvious that $W_{j}(E)$ is determined by
the basin topology, which in turn is determined by the the interaction
energies in the system. Let $W_{j}(E)$ attain its maximum at $E_{j\text{m}}$
and introduce $\Delta_{j\text{m}}E\equiv$ $(\mathcal{E}_{j},E_{j\text{m}}).$
We now introduce the \emph{shifted} PF
\begin{equation}
z_{j}(T)\equiv\underset{E\in\Delta_{j}E}{\sum}W_{j}(E)e^{-\beta(E-\mathcal{E}%
_{j})}\; \label{ShftPF}%
\end{equation}
of the $j$-th basin$.$ Let $E_{j}(T),S_{j}(T)$\ be the \emph{average} basin
energy and entropy, so that $f_{j}(T)\equiv-T\ln z_{j}(T)=E_{j}(T)-\mathcal{E}%
_{j}-TS_{j}(T)$ represents the basin's shifted free energy. Of course, the
conditions determining $E_{j}(T)$ are
\begin{subequations}
\begin{equation}
\left[  \partial S_{j}(E)/\partial E\right]  _{E=E_{j}(T)}=\beta,\text{
}\partial E_{j}(T)/\partial T>0. \label{BasinEquilibrium}%
\end{equation}
Both conditions are always met. We wish to emphasize that $E_{j}(T),S_{j}(T),$
and $f_{j}(T)$ do \emph{not} represent equilibrium quantities yet; the latter
are determined only after $Z(T)$ is evaluated. (If each basin is treated as
representing an independent system in a formal sense, then these quantities do
represent equilibrium values for the particular basin.)

We now group basins, indexed by $j(\lambda)$, into basin classes (BC)
$\mathcal{B}_{\lambda}$, indexed by $\lambda$, so that all basins in a class
have the same BM energy $\mathcal{E}=\mathcal{E}_{\lambda}.$ The basins in a
class do \emph{not} have to be close in the configuration space. The number of
basins in $\mathcal{B}_{\lambda}$ is $N_{\text{BM}}(\mathcal{E}_{\lambda}),$
and the corresponding BM entropy is $S_{\text{BM}}(\mathcal{E}_{\lambda
})\equiv\ln N_{\text{BM}}(\mathcal{E}_{\lambda})$. The introduction of
$N_{\text{BM}}(\mathcal{E}_{\lambda})$ and $S_{\text{BM}}(\mathcal{E}%
_{\lambda})$\ requires that they have a one-to-one relationship with
$\mathcal{E}_{\lambda}.$ Let%

\end{subequations}
\begin{equation}
Z_{\lambda}(T)\equiv\underset{j\in j(\lambda)}{\sum}z_{j}(T),\text{
\ }z_{\lambda}\equiv Z_{\lambda}(T)/N_{\text{BM}}(E_{\lambda}),
\label{BasinClassPF}%
\end{equation}
denote the shifted and the mean (per basin) shifted $\mathcal{B}_{\lambda}%
$-PF, respectively, so that
\begin{subequations}
\begin{align}
Z(T)  &  \equiv\sum_{\lambda}e^{-\beta\mathcal{E}_{\lambda}}Z_{\lambda
}(T)\equiv\sum_{\lambda}e^{-\beta\mathcal{E}_{\lambda}+{}S_{\text{BM}%
}(\mathcal{E}_{\lambda})}z_{\lambda}(T),\,\;\label{ISTotPF1}\\
\mathcal{E}(T)  &  \equiv\sum_{j}\mathcal{E}_{j}e^{-\beta\mathcal{E}_{j}}%
z_{j}(T)/Z(T)\equiv\sum_{\lambda}\mathcal{E}_{\lambda}e^{-\beta\mathcal{E}%
_{\lambda}}Z_{\lambda}(T)/Z(T). \label{ISTotPF2}%
\end{align}
Here, $\overline{\mathcal{E}}=\mathcal{E}(T)$\ represents the equilibrium BM
energy. The equilibrium free energy, entropy and energy are\ $F(T)=-T\ln
Z(T),$ $S(T)=-\partial F/\partial T$ and $E(T)=F(T)+TS(T),$ respectively$.$

At a given $T,$ the probability $\Pr(T\left\vert \mathcal{B}_{\lambda}\right)
$ that the system will probe the BC $\mathcal{B}_{\lambda},$ i.e., any of the
basins in $\mathcal{B}_{\lambda}$ is\
\end{subequations}
\begin{equation}
\Pr(T\left\vert \mathcal{B}_{\lambda}\right)  =e^{-\beta\mathcal{E}_{\lambda}%
}Z_{\lambda}(T)/Z(T). \label{BCProb}%
\end{equation}
On the other hand, the probability that the system will explore a particular
basin $j$ among all the basins is%

\begin{equation}
\Pr(T\left\vert j\right)  =e^{-\beta\mathcal{E}_{\lambda}}z_{j}(T)/Z(T),
\label{BProb}%
\end{equation}
whereas the conditional probability that the system explores the basin $j,$
given that the system is confined to the BC $\mathcal{B}_{\lambda},$ is%
\begin{equation}
p_{j}(T\left\vert \lambda\right)  =z_{j}(T)/Z_{\lambda}(T). \label{B_BCProb}%
\end{equation}
These probability distributions can be used directly to evaluate various
entropies by using the Gibbs definition of the entropy as minus the mean
logarithm of the probability distribution%
\begin{equation}
S=-<\ln w>, \label{SGibbs}%
\end{equation}
where $w$\ is one of the probability distributions above. For example, the
equilibrium entropy $S(T)$ is obtained by considering the distribution in
(\ref{BProb}):%
\begin{equation}
S(T)=-\underset{j}{\sum}\Pr(T\left\vert j\right)  \ln\Pr(T\left\vert j\right)
, \label{SGibbs0}%
\end{equation}
The entropy $\overline{\mathcal{S}}(T)$ due to the number of different
equilibrium BC's, and the entropy $\overline{S}_{\lambda}(T)$ due to basins
within a BC are similarly given by%
\begin{align}
\overline{\mathcal{S}}(T)  &  =-\sum_{\lambda}\Pr(T\left\vert \mathcal{B}%
_{\lambda}\right)  \ln\Pr(T\left\vert \mathcal{B}_{\lambda}\right)
,\label{SGibbs1}\\
\overline{S}_{\lambda}(T)  &  =-\underset{j\in j(\lambda)}{\sum}%
p_{j}(T\left\vert \lambda\right)  \ln p_{j}(T\left\vert \lambda\right)  .
\label{SGibbs2}%
\end{align}
\qquad

It is easy to see that the temperature coefficient $\partial\overline
{\mathcal{E}}/\partial T$ is given by
\begin{equation}
T^{2}\partial\overline{\mathcal{E}}/\partial T=\sum_{j=1}^{N_{\text{B}}}%
\Delta\mathcal{E}_{j}(T)\Delta E_{j}(T)\Pr(T\left\vert j\right)  ,
\label{IdentifierCoefficient}%
\end{equation}
where $\Delta\mathcal{E}_{j}(T)\equiv\mathcal{E}_{j}-\overline{\mathcal{E}},$
$\Delta E_{j}(T)\equiv E_{j}(T)-E(T),$ and $N_{\text{B}}$ is the number of
basins. The derivative is a cross-correlation between two fluctuations,
$\Delta\mathcal{E}_{j}(T)$ and $\Delta E_{j}(T,V)$.\ Since cross-correlations
do not usually have a fixed sign, there is no theoretical reason for
$\overline{\mathcal{E}}$ to be a monotonic increasing function of $T.$ This
does not mean that the temperature coefficient cannot be positive for many
physical systems in a certain temperature range. The same argument also works
for other equilibrium identifier energies.\qquad

The sum over $\lambda$ in (\ref{ISTotPF1},\ref{ISTotPF2}) is over different
basin classes, and the summand is also uniquely determined by each class.
However, the index $\lambda$\ is not a unique labeling. Therefore, we need to
associate with each class a unique thermodynamic class property. We also need
to make sure that its equilibrium value is monotonic in $T$. Thus, our task is
to find a basin quantity that is not only a unique property of each class, but
also determines the summand in (\ref{ISTotPF1},\ref{ISTotPF2}). One of the
simplest choice is to take the basin minima $\mathcal{E}$ as the basin
quantity for the simple reason that each BC has a uniquely BM energy
$\mathcal{E}_{\lambda},$ and it also appears in the summand in (\ref{ISTotPF1}%
,\ref{ISTotPF2}) $\emph{provided}$ we take $z_{\lambda}$ to depend explicitly
on $\mathcal{E}_{\lambda}$ in addition to $T:$ $z_{\lambda}=z_{\lambda
}(\mathcal{E}_{\lambda},T).$ This contradicts the assumption by Goldstein
\cite{Goldstein1}. However, this is a common choice; see \cite{Stillinger}%
.\ If the interest is to study high temperatures near the dynamic crossover,
one might take the LB-energy $\mathcal{E}_{j}^{\text{LB}}$ of the $j$-th basin
as the quantity of interest. In this case, one would group basins not by the
their BM energies, by their LB energies, so that all basins in the
BC\ $\lambda$ have their LB at $\mathcal{E}_{\lambda}^{\text{LB}}.$ Similarly,
one can take the MPB-energy $\mathcal{E}_{j}^{\text{MPB}}$ \ or some other
special energy of the $j$-th basin as the quantity of interest. The analysis
in all these cases would be carried out in the same manner as the analysis we
carry out here for the BM picture, and would not be pursued further except for
a few remarks later.

\textbf{Common Assumption. }The assumption $z_{\lambda}=z_{\lambda
}(\mathcal{E}_{\lambda},T)$ as a function of \emph{two different variables
}$T,$ and $\mathcal{E}_{\lambda}$ makes the summand an explicit function of
$\mathcal{E}_{\lambda}.$\ Note that $\mathcal{E}_{\lambda}$\ in the summand in
(\ref{ISTotPF1},\ref{ISTotPF2}) is a temperature-independent quantity. Let us
follow the consequence of this \emph{assumption.} Let the mean basin free
energy resulting from the mean basin PF $z_{\lambda}$ for the basin class
$\mathcal{B}_{\lambda},$ and the mean basin entropy be $f(\mathcal{E}%
_{\lambda},T)=-T\ln z_{\lambda}(\mathcal{E}_{\lambda},T),$ and $S(\mathcal{E}%
_{\lambda},T)=-[\partial f(\mathcal{E}_{\lambda},T)/\partial T\mathcal{]}%
_{\mathcal{E}_{\lambda}},$ respectively.\ The mean basin free energy
$f(\mathcal{E}_{\lambda},T)=E_{\lambda}(T)-\mathcal{E}_{\lambda}-T$
$S(\mathcal{E}_{\lambda},T)$ is obviously obtained by minimizing the free
energy function $f_{\lambda}(\mathcal{E}_{\lambda},E,T)=E-\mathcal{E}%
_{\lambda}-T$ $S_{\lambda}(\mathcal{E}_{\lambda},E)$ with respect to $E$ at
constant $T$ and $\mathcal{E}_{\lambda}.$ Here, $S_{\lambda}(\mathcal{E}%
_{\lambda},E)\equiv S_{\lambda}(E)-S_{\text{BM}}(\mathcal{E}_{\lambda}),$ and
$\exp[S_{\lambda}(E)]$\ represents the number of configurations of energy $E$
that belong to $\mathcal{B}_{\lambda}.$ The condition for the minimum is
$[\partial S_{\lambda}(\mathcal{E}_{\lambda},E)/\partial E]_{\mathcal{E}%
_{\lambda}}=\beta$ at $E=E_{\lambda}(T),$ which determines $f(\mathcal{E}%
_{\lambda},T)=E_{\lambda}(T)-\mathcal{E}_{\lambda}-T$ $S(\mathcal{E}_{\lambda
},T);$ here, $S(\mathcal{E}_{\lambda},T)=S_{\lambda}[\mathcal{E}_{\lambda
},E_{\lambda}(T)].$

The sum over the index $\lambda$ is now replaced by the BM variable
$\mathcal{E=E}_{\lambda}.$\ Because of the assumed $\mathcal{E}$-dependence,
the general summand in $(\ref{ISTotPF1},\ref{ISTotPF2})$ becomes an explicit
function of $\mathcal{E},$\ and we can minimize the corresponding free energy
function $F_{\text{B}}(\mathcal{E},T)\equiv\mathcal{E}+f(\mathcal{E}%
,T)-TS_{\text{BM}}(\mathcal{E})$ with respect to $\mathcal{E}$ at
fixed\emph{\ }$T$ to determine $Z(T)$ for a macroscopic system.\ The minimum
of $F_{\text{B}}(\mathcal{E},T)$ is located by the condition $[\partial
F_{\text{B}}(\mathcal{E},T)/\partial\mathcal{E}]_{T}=0.$ This condition is
satisfied at the equilibrium BM-energy $\overline{\mathcal{E}}%
\mathcal{=\mathcal{E}(}T\mathcal{)=E}(T),$ see (\ref{ISTotPF2}). It is also
equivalently given by the solution of
\begin{equation}
\lbrack\partial S_{\text{BM}}(\mathcal{E})/\partial\mathcal{E}]_{\mathcal{E}%
=\overline{\mathcal{E}}}=\beta\{1+[(\partial f(\mathcal{E},T)/\partial
\mathcal{E)}_{T}]_{\mathcal{E}=\overline{\mathcal{E}}}\}, \label{ISTMaxCon}%
\end{equation}
and determines the equilibrium free energy $F(T)\equiv$ $F_{\text{B}%
}(\overline{\mathcal{E}},T)$ and the BM-entropy $S_{\text{BM}}(T)\equiv$
$S_{\text{BM}}(\overline{\mathcal{E}}).$ The solution of (\ref{ISTMaxCon})
yields $\overline{\mathcal{E}}\mathcal{=\mathcal{E}(}T\mathcal{)}$ as a
function of a single variable $T.$ The equilibrium mean basin free energy and
entropy are $f_{\text{b}}(T)=f(\overline{\mathcal{E}},T)$ and $S_{\text{b}%
}(T)=S(\overline{\mathcal{E}},T),$ respectively. It is easy to see that the
form of the equilibrium free energy $F_{\text{B}}(T)=f(\overline{\mathcal{E}%
},T)+\mathcal{E(}T\mathcal{)}-TS_{\text{BM}}(\overline{\mathcal{E}})$\ is the
same as the free energy obtained by Goldstein in (\ref{GoldsteinPart}), except
that the equations determining the equilibrium BM-energy are different;
compare (\ref{Goldstein1}) and (\ref{ISTMaxCon}). The two conditions become
identical if $f$ is taken to be independent of $\mathcal{E},$ as was assumed
by Goldstein.

Let us assume that $\overline{\mathcal{E}}=\mathcal{E(}T\mathcal{)}$\ is a
\emph{monotonic} function of $T$, so that we can invert it to express
$T=T(\overline{\mathcal{E}}),$ which can be used to express all $T$-dependent
quantities like $f_{\text{b}}(T),F(T),S_{\text{BM}}(T),$ etc. as functions of
the equilibrium BM-energy $\overline{\mathcal{E}}.$ We can also express
$f(\mathcal{E}_{\lambda},T)$\ as a function of $\mathcal{E}_{\lambda}$ and
$\overline{\mathcal{E}}.$ All this means is that $T$ and $\overline
{\mathcal{E}}$\ are equivalent. The equilibrium free energy $\overline
{f_{\text{b}}}(\overline{\mathcal{E}})\equiv f_{\text{b}}[T(\overline
{\mathcal{E}})]$ is a single-variable function of $\overline{\mathcal{E}}$ and
says nothing uniquely about the $\mathcal{E}$-dependence in the two-variable
function $f(\mathcal{E},T).$ This is most clearly seen in the Goldstein's
scenario. While $f_{\text{b}}(T)$ in his analysis is independent of
$\mathcal{E},$ we can still express $f_{\text{b}}(T)$ as $\overline
{f_{\text{b}}}(\overline{\mathcal{E}}).$ To conclude that $\overline
{f_{\text{b}}}(\overline{\mathcal{E}})$ implies that $f_{\text{b}}(T)$ is a
two variable function having an additional $\mathcal{E}$-dependence would be a
wrong conclusion. A similar inversion can be carried out in terms of the
average LB-energy $\mathcal{E}_{\lambda}^{\text{LB}}(T)$ so that all
quantities that depend on $T$ can be expressed as a function of $\mathcal{E}%
_{\lambda}^{\text{LB}}(T),$ which suggests that there is nothing unique in the
choice of BM energies as the thermodynamic quantity to replace the sum over
$\lambda.$

In the above analysis, we have only considered one ($[\partial F_{\text{B}%
}(\mathcal{E},T)/\partial\mathcal{E}]_{T}=0$) of the two conditions needed for
the free energy minimization. The other condition is that the curvature of the
free energy at the minimum be positive:$[\partial^{2}F_{\text{B}}%
(\mathcal{E},T)/\partial\mathcal{E}^{2}]_{T}>0.$ This issue has been
investigated elsewhere \cite{Gujrati}, and will not be pursued here.

\textbf{Current Analysis}. We now proceed to prove our central result that
$\mathcal{E}_{\lambda}$ (or any other basin identifiers) cannot be the choice
of the thermodynamic quantity to replace the sum over $\lambda$ for two
different reasons that will be discussed below$.$ We observe that $W(E)$ is a
\emph{sum} over various basins containing the energy $E$: $W(E)\equiv
\sum_{j:E\in\Delta_{j}E}W_{j}(E).$ We also observe that even for basins in a
single BC $\mathcal{B}_{_{\lambda}}$, $z_{j}(T)$ need not be equal.

1.\qquad\textbf{Evaluating} $\mathit{z}_{j}\mathbf{(\mathit{T})}$\textbf{. }We
first prove one of the assumptions by Goldstein \cite{Goldstein1}
that\ $z_{j}$ \emph{cannot} depend explicitly on the BM energy $\mathcal{E}%
_{j}$, see (\ref{ShftPF}), though it most certainly depends on the shape of
the basin, i.e. on $j$. For example, the basin curvature and not
$\mathcal{E}_{j}$ determines the vibrational frequencies (including
anharmonicity) which, in turn, determine the free energy $f_{j}(T)\equiv-T\ln
z_{j}(T)$ in the harmonic approximation$.$ The latter is measured with respect
to $\mathcal{E}_{j},$ so it must be independent of $\mathcal{E}_{j}.$

To be convinced that this is most certainly correct, consider two basins $j=$1
and 2 that are identical in all respects (shape and topology), but have
different BM energies, which we assume satisfy $\mathcal{E}_{1}<$%
\ $\mathcal{E}_{2}.$ Let us introduce $\delta_{j}E=E-\mathcal{E}_{j}.$ Let
$W_{j}^{\prime}(\delta_{j}E),$ $S_{j}^{\prime}(\delta_{j}E)=\ln W_{j}^{\prime
}(\delta_{j}E)$\ denote the number of , and entropy due to configurations of
energy at a height $\delta_{j}E$ above the basin minimum $\mathcal{E}_{j}$ in
the $j$-th basin$.$ It is obvious that%
\begin{equation}
W_{j}^{\prime}(\delta_{j}E)=W_{j}(E). \label{EShift}%
\end{equation}
Now, for $\delta_{1}E=\delta_{2}E=\delta E,$ it is clear that $S_{1}^{\prime
}(\delta E)=S_{2}^{\prime}(\delta E)$ for the two identical basins, from which
it immediately follows that $\partial S_{1}^{\prime}(\delta E)/\partial\delta
E=\partial S_{2}^{\prime}(\delta E)/\partial\delta E=\beta.$ The free energies
of the two basins are given in terms of the same average energy height $\delta
E(T)=E_{j}(T)-\mathcal{E}_{j},$ where $E_{j}(T)$\ is the average basin energy
in the $j$-th basin, and the same entropy $S_{1}^{\prime}[\delta
E(T)]=S_{2}^{\prime}[\delta E(T)]:$
\[
f_{j}(T)=\delta E(T)-TS_{j}^{\prime}[\delta E(T)]=E_{j}(T)-\mathcal{E}%
_{j}-TS_{j}(T),
\]
and are equal, as expected, after we use (\ref{EShift}). Here, $S_{j}(T)$\ is
the canonical basin entropy in the $j$-th basin. We can also check the
veracity of the free energy equality by directly evaluating the shifted PF\ in
(\ref{ShftPF}). We will not do this as it is trivial. The extension to many
topologically identical basins is trivial. Thus, we conclude that the free
energies $f_{j}(T)$ do not depend on the BM energies $\mathcal{E}_{j}$.

Indeed, a very general but direct proof can be given that the basin PF
$z_{j}(T)$ cannot depend on its BM energy $\mathcal{E}_{j}.$ Thus, the
corresponding basin entropy $S_{j}(T)$ is also independent of $\mathcal{E}%
_{j}.$\ For this, let us shift all energies $E\rightarrow E^{^{\prime}}\equiv
E-C$ by some constant $C$ in (\ref{ShftPF})$.$ The number $W_{j}(E)$ of
states, all having the same energy $E,$ remains unchanged under the shift by
$C$: $W_{j}(E)$ $\rightarrow W_{j}^{\prime}(E^{^{\prime}})=W_{j}(E);$ compare
with (\ref{EShift})$.$ Thus, $z_{j}$ transforms under the shift as
$z_{j}(T)\rightarrow\sum_{E^{^{\prime}}}W_{j}^{^{\prime}}(E^{^{\prime}%
})e^{-\beta(E^{^{\prime}}-E_{j}^{^{\prime}})},$ see (\ref{ShftPF}), and
remains unchanged for any arbitrary $C,$ including $C=\mathcal{E}%
_{j},\mathcal{E}_{j}^{\text{B}},$ or $\mathcal{E}_{j}^{\text{MPB}}$
corresponding to the $j$-th basin.

This central result requires adopting a different approach than the above for
the PF-evaluation, which we present below.

For a macroscopic system, $z_{j}$ is determined by\ the maximum summand in
(\ref{ShftPF}) corresponding to $E=\overline{E_{j}}=E_{j}(T)\in\Delta_{j}E.$
In terms of the free energy function $\varphi_{j}(E,T)\equiv E-TS_{j}(E),$ the
energy $\overline{E_{j}}$\ is determined by $(\partial\varphi_{j}/\partial
E)_{T}=0.$\ As usual, this condition yields
\begin{equation}
(\partial S_{j}(E)/\partial E)_{E=\overline{E_{j}}}=\beta,\;\partial
E_{j}(T)/\partial T\;>0,\;\overline{E_{j}}\in\Delta_{j\text{m}}E;
\label{EqScond}%
\end{equation}
the restriction on the allowed values of $\overline{E_{j}}$\ ensures that we
are only considering non-negative $\beta.$ Over this temperature range,
$S_{j}(T)=S_{j}(\overline{E_{j}})$\ and $E_{j}(T)$\ are monotonic increasing.
On the other hand, the shifted free energy of the basin is $f_{j}%
(T)=E_{j}(T)-\mathcal{E}_{j}-TS_{j}(T)$ is monotonic decreasing since
$S_{j}(T)=-[\partial f_{j}(T)/\partial T]_{\mathcal{E}_{j}}\geq0.$ \ Since
there is no $\mathcal{E}_{j}$-dependence in $f_{j}(T),$ as we have argued
above, there is no need to show that the derivative $\partial f_{j}%
(T)/\partial T$ is at fixed $\mathcal{E}_{j}$.

The energy landscape is topologically very complex, with various basins very
different from each other, even if they have their minima at the same energy.
Thus, $E_{j}(T)$ in different basins at the same temperature $T$ will be
usually different not only from each other, but also from $E(T)$ that appears
in (\ref{EquilibriumE}). In this sense, all basins are independent at this
step of the analysis as they all do not represent equilibrium configurations.
By the last step of the analysis, only equilibrium basins will survive for
which $E_{j}(T)=E(T).$

2.\qquad\textbf{Evaluating}$\ \mathit{Z}_{\lambda}\mathbf{(\mathit{T})}%
$\textbf{.} As shown above, $z_{j}(T)$ is not a function of $\mathcal{E}%
_{\lambda}.$\ We now need to replace the summation over $j$ in
(\ref{BasinClassPF}) by a suitable thermodynamic quantity. There are various
choices for the sum in (\ref{BasinClassPF}) like the basin free energy
$f_{j}(T),$ the basin energy $E_{j}(T),$ or the basin entropy $S_{j}(T),$ all
equally good$.$ All these quantities are monotonic in $T.$ We will choose
$f_{j}(T)$ in the following to be the thermodynamic quantity to replace the
sum over the index $j.$\ As the basins in $\mathcal{B}_{_{\lambda}}$ do not
all have the same free energy, we classify each basin in (\ref{BasinClassPF})
according to its free energy $f$ at a given temperature $T$. Let $N_{\lambda
}(f,T)$\ denote the number of basins of free energy $f$ at a given $T$ in
$\mathcal{B}_{\lambda},$ and let $S_{\lambda}(f,T)\equiv\ln N_{\lambda}(f,T)$
be the corresponding entropy. For a macroscopic system at a given fixed $T$,
$Z_{\lambda}$ is dominated by the basins in $\mathcal{B}_{\lambda}$ for which
$F_{\lambda}(f,T)\equiv f(T)-TS_{\lambda}(f,T)$ is minimum as a function of
$f$ at\emph{\ fixed} $T$, the conditions for which are $(\partial F_{\lambda
}/\partial f)_{T}=0,$ and $(\partial^{2}F_{\lambda}/\partial f^{2})_{T}>0.$
The resulting entropy and free energy at the minimum $(f=\overline{f}%
_{\lambda})$ are denoted by $\overline{S}_{\lambda}(T)=S_{\lambda}%
(\overline{f}_{\lambda},T),$ and $\overline{F}_{\lambda}(T)=\overline
{f}_{\lambda}-T\overline{S}_{\lambda},$ respectively. One of the conditions
for the minimum is given by%
\begin{equation}
\lbrack(\partial S_{\lambda}/\partial f)_{T\text{ }}]_{f=\overline{f}%
_{\lambda}}=\beta.
\end{equation}
From (\ref{B_BCProb}), we find the conditional probability $p_{j}(T\left\vert
\lambda\right)  =\exp[-\beta(f_{j}-\overline{F}_{\lambda})].$ Using this in
(\ref{SGibbs2}), we can evaluate the entropy due to the number of basins in
$\mathcal{B}_{\lambda}.$ It is easy to see that this entropy is precisely
$\overline{S}_{\lambda}(T),$ as expected.

Note that $\mathcal{E}_{\lambda}$ is independent of $T.$ Using once more the
shift argument $E\rightarrow E^{^{\prime}}\equiv E-C,$ and the invariance of
$z_{j}$ so that the sum in (\ref{BasinClassPF}) contains the same number of
basins $N_{\lambda}(f,T),$ we similarly conclude that $Z_{\lambda}(T)$ also
remains unchanged for any arbitrary $C,$ including $C=\mathcal{E}_{\lambda}.$
Thus, the shifted BC free energy $\overline{F}_{\lambda}(T)$ is
\emph{independent }of $\mathcal{E}_{\lambda}$, although it most certainly
depend on the BC $\mathcal{B}_{\lambda}.$ In other words, it \emph{cannot} be
written as a two-variable function $\overline{F}_{\lambda}(\mathcal{E}%
_{\lambda},T).$ The corresponding entropy $\overline{S}_{\lambda}(T)$ is also
\emph{independent }of $\mathcal{E}_{\lambda}.$

Even for the singular case, when $z_{j}(T)$ for each basin in the BC
$\mathcal{B}_{\lambda}$ is the same, all basins [their number being
$N_{\text{BM}}(\mathcal{E}_{\lambda})$] contribute. In this case, $N_{\lambda
}(f,T)=N_{\text{BM}}(\mathcal{E}_{\lambda})$ and the above formulation
continues to work.

3.\qquad\textbf{Evaluating}$\ \mathit{Z}\mathbf{(\mathit{T})}$. Using the
evaluated $Z_{\lambda}$ in $Z$, we find
\begin{equation}
Z=\underset{\lambda}{\sum}e^{-\beta\lbrack\mathcal{E}_{\lambda}+\overline
{F}_{\lambda}(T)]}. \label{Total}%
\end{equation}
Since we have already seen that $\overline{F}_{\lambda}(T)$ is independent of
$\mathcal{E}_{\lambda}$, the summation over $\lambda$ \emph{cannot} be
replaced by a summation over $\mathcal{E}_{\lambda}.$ We need to look for a BC
quantity that can be uniquely related to the index $\lambda$ and which also
uniquely determines the summand. For this we proceed in a \emph{standard}
manner as follows$.$ Let $\mathcal{N}(\mathcal{F},T)$ denote the number of
BC's at a given $T$\ with the same $\mathcal{F\equiv E}_{\lambda}+\overline
{F}_{\lambda}(T),$ where $\mathcal{F}$ is the \emph{unshifted} BC free energy
now with respect to the zero of $E.$ Since $\mathcal{F}$ for different basin
classes are now measured from the same common point $E=0,$ it is clear that if
we consider all those BC's that have the same free energy $\mathcal{F}$ at
some $T,$ then this $\mathcal{F}$\ no longer\emph{ depends }on the individual
BM energy $\mathcal{E}_{\lambda}$. The PF $Z$ is dominated by the BC's for
which $\mathcal{F}-T\mathcal{S}$ is minimum over $\mathcal{F}$; here
$\mathcal{S}(\mathcal{F},T)\equiv\ln\mathcal{N}(\mathcal{F},T).$ One of the
conditions for this minimum at $\overline{\mathcal{F}}$ is
\begin{equation}
\lbrack(\partial\mathcal{S}/\partial\mathcal{F)}_{T}]_{\mathcal{F=}%
\overline{\mathcal{F}}}=\beta,\; \label{ISC}%
\end{equation}
as expected. Hence, we finally conclude that the free energy of the system is
given by
\begin{equation}
F(T)\equiv-T\ln Z\equiv\overline{\mathcal{F}}-T\overline{\mathcal{S}};
\label{FinalFreeEnergy}%
\end{equation}
here, $\overline{\mathcal{S}}(T)=\mathcal{S}(\overline{\mathcal{F}},T).$ It
should be obvious at this point that the dominant contribution in
(\ref{Total}) will usually mix BC's with different $\mathcal{E}_{\lambda}.$

The equilibrium free energy is the sum of three terms: $F(T)=\varphi
_{\text{b}}(T)-T\overline{S}_{\text{BC}}(T)-T\overline{\mathcal{S}}(T),$ where
$\overline{S}_{\text{BC}}$ denotes the equilibrium basin class entropy and
$\varphi_{\text{b}}(T)$ the equilibrium basin free energy. Expressing
$\varphi_{\text{b}}(T)=E_{\text{b}}(T)-TS_{\text{b}}(T),$ we immediately see
that
\begin{equation}
S(T)=S_{\text{b}}(T)+\overline{S}_{\text{BC}}(T)+\overline{\mathcal{S}}(T).
\label{EntropyPartition}%
\end{equation}

Introducing $\mathcal{F}_{\lambda}\mathcal{\equiv E}_{\lambda}+\overline
{F}_{\lambda}(T),$ we can express the probability $\Pr(T\left\vert
\mathcal{B}_{\lambda}\right)  ,$ see (\ref{BCProb}), that the system will
explore $\mathcal{B}_{\lambda}$ in terms of $\mathcal{F}_{\lambda},$ \
\begin{equation}
\Pr(T\left\vert \mathcal{B}_{\lambda}\right)  =e^{-\beta\lbrack\mathcal{F}%
_{\lambda}-F]}. \label{ProbBC}%
\end{equation}
It is easy to check that using the above probability in (\ref{SGibbs1}), we
obtain the above entropy $\overline{\mathcal{S}}(T).$ In terms of the above
probability in (\ref{ProbBC}), the equilibrium BM energy, see (\ref{ISTotPF2}%
), is expressed as follows:
\begin{equation}
\mathcal{E}(T)\equiv\sum_{\lambda}\mathcal{E}_{\lambda}\Pr(T\left\vert
\mathcal{B}_{\lambda}\right)  . \label{FinalBMEnergy}%
\end{equation}
Thus, unless $\overline{\mathcal{S}}(T)=0,$ so that there is only one BC, the
equilibrium BM energy $\mathcal{E}(T)$ is not given by the BM energy of a
single BC. In other word, the system will usually explore many basin classes
of \emph{different} BM energies, unless $\overline{\mathcal{S}}(T)=0.$ it is
expected that $\overline{\mathcal{S}}(T)$ vanishes at low temperatures. Thus,
the use of BM's makes perfect sense at low temperatures as Goldstein had
argued \cite{Goldstein,Goldstein1}. At higher temperatures, many BC's with
different $\mathcal{E}_{\lambda}$ are involved in (\ref{FinalBMEnergy}).

The last prediction is different from the BM approach in which the equilibrium
basins all have the same average BM energy $\mathcal{E}(T).$ Let us try to
understand this point carefully. Let us group the BC's according to their free
energy $\mathcal{F}.$ The number of BC's in a group $\mathcal{G}%
(\mathcal{F},T)$ corresponding to a given $\mathcal{F}$\ is $\mathcal{N}%
(\mathcal{F},T).$\ Let
\begin{equation}
\mathcal{E(F},T)=\sum_{\lambda\in\mathcal{G}(\mathcal{F},T)}\mathcal{E}%
_{\lambda}/\mathcal{N}(\mathcal{F},T) \label{FinalBMEnergy0}%
\end{equation}
denote the mean BM group energy in the group $\mathcal{G}(\mathcal{F},T).$
Then,%
\begin{equation}
\mathcal{E(}T)=\sum_{\mathcal{G}(\mathcal{F},T)}\mathcal{E(F},T)e^{\mathcal{S}%
(\mathcal{F},T)-\beta\lbrack\mathcal{F}-F]}. \label{FinalBMEnergy1}%
\end{equation}
For a macroscopic system, the dominant term in the above sum
(\ref{FinalBMEnergy1}) corresponds to $\mathcal{F=}\overline{\mathcal{F}},$ so
that
\begin{equation}
\mathcal{E(}T)=\mathcal{E(}\overline{\mathcal{F}},T). \label{FinalBMEnergy2}%
\end{equation}
Unless $\mathcal{N}(\overline{\mathcal{F}},T)=1,$ the sum in
(\ref{FinalBMEnergy0}) contains many BC's, each with different $\mathcal{E}%
_{\lambda}.$ Since the summand in (\ref{FinalBMEnergy0}) does not have any
coeeficient similar to $W(E)$ in (\ref{StandardPF}), there is no reason for
$\mathcal{E}_{\lambda}^{\prime}$s to peak around the mean $\mathcal{E(F},T).$
Thus, the equilibrium BM energy $\mathcal{E(}T)$ represents a mean group BM
energy over many different BM energies $\mathcal{E}_{\lambda}.$

From (\ref{FinalBMEnergy2}), we conclude that $\mathcal{E(}T)$ should be
properly thought of as the equilibrium value of a two-variable function
$\mathcal{E(F},T).$ In this form, the change in $\mathcal{E(}T)$ has two
different contributions. One contribution is due to the intrabasin temperature
change $dT$ at constant $\mathcal{F}$, and the other contribution comes from
basin changes caused by free energy change in $d\mathcal{F}$ at constant $T.$
There is no reason for either contribution to vanish in general.

It is also easy to evaluate equilibrium values $\mathcal{E}^{\text{B}%
}\mathcal{(}T),$ or $\mathcal{E}^{\text{MPB}}\mathcal{(}T)$ using our
formalism. Using the probability $\Pr(T\left\vert j\right)  ,$ see
(\ref{BProb}), which is given by%
\begin{equation}
\Pr(T\left\vert j\right)  =e^{-\beta\lbrack f_{j}(T)+\mathcal{E}_{j}-F(T)]},
\label{BProb1}%
\end{equation}
we can calculate the above equilibrium energies as follows:%
\begin{equation}
\mathcal{E}^{\text{LB}}\mathcal{(}T)=\sum_{j}\mathcal{E}_{j}^{\text{LB}}%
\Pr(T\left\vert j\right)  ,\mathcal{E}^{\text{MPB}}\mathcal{(}T)=\sum
_{j}\mathcal{E}_{j}^{\text{MPB}}\Pr(T\left\vert j\right)  .
\label{FinalBHPBEnergy}%
\end{equation}
Following a similar analysis that we have carried out here using BM as the
point of interest, we can derive a similar relation like (\ref{FinalBMEnergy2}%
) for the above two equilibrium energies. We will not do this here. By
comparing the equilibrium energy $E(T)$ with the equilibrium values in
(\ref{FinalBHPBEnergy}), we can draw conclusions about the importance of
various basin identifiers at a given temperature.

The landscape analysis developed here does not involve the equilibrium BM
energy $\overline{\mathcal{E}},$ because $\mathcal{E}_{j}$ are not the natural
energies that appear in the PF. Despite this, it is possible to express
thermodynamic quantities in terms of $\overline{\mathcal{E}}$ under certain
conditions that we elucidate below. The equilibrium basins at a given
temperature form a subset of all basins and determine the equilibrium
thermodynamics. If $\mathcal{E}(T)$ is \emph{monotonic} in $T,$ we can invert
the relation $\mathcal{E}(T)$ as before$:T=T(\overline{\mathcal{E}}).$ Thus,
we can express any function of $T$ as a function of $\overline{\mathcal{E}}.$
For example, the equilibrium basin free energy $f_{\text{b}}(T)$ can be
expressed as a function of $\overline{\mathcal{E}}=\mathcal{E}(T).$\ This is
one way to express the equilibrium free energy as a function of $\mathcal{E}%
(T).$\ There are other ways to do this, as we will discuss below.

We can use $E(T)$ at a given $T$ to select only the equilibrium basins whose
average energy is $E(T)$. This is precisely what is done when equilibrium
configurations are generated in simulations carried out at a fixed $T$ without
any consideration of the basins \cite{Wales}. From these equilibrium
configurations, equilibrium basins can be identified. These basins also appear
in (\ref{FinalBMEnergy}) and determine $\overline{\mathcal{E}}.$ But
simulations can also provide information about whether there are many
different BC's that contribute to $\overline{\mathcal{E}}.$ (Unfortunately, to
the best of our knowledge, simulations carried on finite-size systems have not
been used to unequivocably answer whether the spread in the allowed values of
$\mathcal{E}$ are thermodynamically significant are not.) The equilibrium free
energy $f(T)$ of these basins will depend on the average topology of the
basins. For example, in the harmonic approximation, $f(T)$ will be determined
by the curvature (given by the set $\nu$ of vibrational frequencies) at the
BM, and should be expressed as a function $f(\nu,T)$. These equilibrium
frequencies will change with $T.$ Being a function of $T,$ they can also be
expressed as a function of $\overline{\mathcal{E}}$ as discussed above. Thus,
the equilibrium basin free energy can be expressed as a function
$f(\overline{\mathcal{E}},T).$ This provides us with another way to express
$f(T)$\ as a function of $\overline{\mathcal{E}}.$ Such a dependence on
$\overline{\mathcal{E}}$\ of the equilibrium basin free energy should
\emph{not} be confused with any explicit $\mathcal{E}_{j}$-dependence of
$z_{j}(T)$ in (\ref{ShftPF}), or $z_{\lambda}(T)$ in (\ref{ISTotPF1}), since
such a dependence has been shown not to exist. The free energy in
(\ref{ISTMaxCon}) is \emph{not} this equilibrium free energy in its new form.
Moreover, simulations usually do not generate basins other than the
equilibrium basins. Thus, they provide no information about the free energy
$f$ that appears in (\ref{ISTMaxCon}).

It should be clear that depending on the complexity of the basin potential,
there are many ways to express $f(T)$\ as a function of $\overline
{\mathcal{E}}.$ For example, the anharmonicity, to be represented in short by
a set of parameters $a,$ of the equilibrium basin will also depend on $T.$
Now, we may express both $\nu$ and $a$ as a function of $\overline
{\mathcal{E}},$ or decide only to express $\nu$ as a function of
$\overline{\mathcal{E}},$ and leave $a$ as a function of $T.$ This will give
us two different forms of the equilibrium basin free energy, with both
carrying the same information as a function of $T,$ but not as a function of
$\overline{\mathcal{E}}.$\ Thus, such representations are obviously not
unique. This is because of the independence of the basin free energy and
$\mathcal{E}.$ On the other hand, if a quantity has an explicit $\mathcal{E}%
$-dependence, we can immediately obtain its equilibrium value at a given $T$
by replacing $\mathcal{E}$ by $\overline{\mathcal{E}}.$ For example,
$S_{\text{BM}}(\mathcal{E})$ can be expressed in a unique way as a function of
$T$ by identifying $S_{\text{BM}}(T)=S_{\text{BM}}(\overline{\mathcal{E}}).$

The above argument is easily extended to expressing thermodynamic quantities
in terms of $\mathcal{E}^{\text{LB}}\mathcal{(}T),\mathcal{E}^{\text{MPB}%
}\mathcal{(}T),$ or any other particular average energy related to the basins.
There is nothing unique about the BM energy in the landscape, except possibly
at low temperatures where the equilibrium free energy is close to
$\mathcal{E}(T).$

In summary, we have made a distinction between basins and equilibrium basins
and quantities related to them, which is usually not done. With this
distinction in place, we have shown that the shifted basin free energy is
independent of particular basin energies like that of BM, LB, MPB, etc. It is
a function only of $T.$ Thus, the equilibrium basin free energy is also a
function of $T,$ which is consistent with the first assumption of Goldstein.
It is possible to express the equilibrium basin energy as a function of
$\mathcal{E}(T)$,$\mathcal{E}^{\text{LB}}\mathcal{(}T)$,$\mathcal{E}%
^{\text{MPB}}\mathcal{(}T)$ etc. in presumably many ways, as discussed above,
but only if $\mathcal{E}(T)$,$\mathcal{E}^{\text{LB}}\mathcal{(}%
T)$,$\mathcal{E}^{\text{MPB}}\mathcal{(}T)$ etc. are monotonic functions of
$T$. The unshifted equilibrium BC free energy $\mathcal{F}$ that contributes
to (\ref{FinalFreeEnergy}) is also independent of the BM energy $\mathcal{E}$
by construction. This is consistent with the second assumption of Goldstein,
albeit in a modified form. Thus, the configurational states that are to be
associated with the entire topology of the basins cannot be characterized
simply by their BM energies $\mathcal{E}$, LB energies $\mathcal{E}%
^{\text{LB}},$ MPB energy $\mathcal{E}^{\text{MPB}}$ or some other particular
basin energy $;$ however, the basin free energy can be used to specify the
configurational states. The current analysis deals only with the basin free
energies. We have found that there is not one average BC of a given average BM
energy $\mathcal{E}(T),$ but many different BC's are explored unless
$\overline{\mathcal{S}}(T)=0.$

It is our pleasure to thank Andrea Corsi for his comments on the work.

\end{document}